# MODELLING ERRORS IN X-RAY FLUOROSCOPIC IMAGING SYSTEMS USING PHOTOGRAMMETRIC BUNDLE ADJUSTMENT WITH A DATA-DRIVEN SELF-CALIBRATION APPROACH

J. C. K. Chow [1,2,*], D. D. Lichti [3], K. D. Ang [2,4], G. Kuntze [5], G. Sharma [5], J. Ronsky [5]

[1] Department of Medicine, Cumming School of Medicine, University of Calgary, Calgary, Alberta, Canada – jckchow@ucalgary.ca
[2] Department of Research and Development, Vusion Technologies, Calgary, Alberta, Canada
[3] Department of Geomatics Engineering, Schulich School of Engineering, University of Calgary, Calgary, Alberta, Canada – ddlichti@ucalgary.ca
[4] Department of Computer Science, Faculty of Science, University of Calgary, Calgary, Alberta, Canada – kdang@ucalgary.ca
[5] Department of Mechanical and Manufacturing Engineering, Schulich School of Engineering, University of Calgary, Calgary, Alberta, Canada – (gkuntze, gbsharma, jlronsky)@ucalgary.ca

**Commission I, WG I/9**

**ABSTRACT:**

X-ray imaging is a fundamental tool of routine clinical diagnosis. Fluoroscopic imaging can further acquire X-ray images at video frame rates, thus enabling non-invasive in-vivo motion studies of joints, gastrointestinal tract, etc. For both the qualitative and quantitative analysis of static and dynamic X-ray images, the data should be free of systematic biases. Besides precise fabrication of hardware, software-based calibration solutions are commonly used for modelling the distortions. In this primary research study, a robust photogrammetric bundle adjustment was used to model the projective geometry of two fluoroscopic X-ray imaging systems. However, instead of relying on an expert photogrammetrist's knowledge and judgement to decide on a parametric model for describing the systematic errors, a self-tuning data-driven approach is used to model the complex non-linear distortion profile of the sensors. Quality control from the experiment showed that 0.06 mm to 0.09 mm 3D reconstruction accuracy was achievable post-calibration using merely 15 X-ray images. As part of the bundle adjustment, the location of the virtual fluoroscopic system relative to the target field can also be spatially resected with an RMSE between 3.10 mm and 3.31 mm.

## 1. INTRODUCTION

X-ray fluoroscopy is a valuable diagnostic imaging modality in gastroenterology, radiology, orthopaedics and many other medical specialities. For example, a barium swallow under a fluoroscope is the key confirmatory tool if a physician suspects achalasia in patients presenting with progressive esophageal dysphagia pertaining to both solids and liquids. Contrast fluoroscopy allows for the study of gastrointestinal tract motility by measuring any obstructions and tracking the velocity of a fluid bolus. In order to accurately measure the diameter of the gastrointestinal tract and/or quantify the velocity of particles, the fluoroscopic imaging system needs to be calibrated to minimize the systematic errors. This article proposes a scalable data-driven approach to model the distortions experienced in X-ray fluoroscopy. Quality control is performed by comparing the reconstructed object space positions and the relative sensor positions to a reference solution.

## 2. BACKGROUND

Direct Linear Transformation (DLT) type methods are often used for calibrating fluoroscopic imaging systems because of their computational simplicity (You et al., 2001). However, it assumes that the systematic errors in every image frame are independent of the other image frames acquired by the same sensor. In a well-constructed and stable system, the systematic errors should be relatively constant with a very slow drift (in other words, it has a high temporal stability). Therefore, it is reasonable to hypothesize that a set of X-ray images captured by the same sensor will share the same distortion profile. By considering a set of images concurrently in the distortion modelling process, the accuracy and robustness of the calibration can be improved.

To estimate a unique set of calibration parameters for a dual fluoroscopic X-ray imaging system, Lichti et al. (2015) demonstrated the use of the photogrammetric bundle adjustment method to combine image space information extracted from 300 images. The authors reported achieving up to 71% improvement in 3D reconstruction accuracy after calibration. This method was further extended in Al-Durgham et al. (2016) where a semi-automatic target extraction and matching function was added to make the entire calibration process more efficient and user-friendly. However, up until now an expert photogrammetrist was still required to study the residuals graphically and perform statistical analysis to determine the appropriate model complexity. In this paper, a data-driven approach is proposed to help make the calibration procedure operator-independent by automatically selecting the most appropriate distortion profile based on the input data during bundle adjustment.

## 3. MATHEMATICAL MODEL

The fundamental basis of the proposed calibration method is that the physical arrangement of the high-speed camera, X-ray source, and image intensifier found in a fluoroscopic imaging system can be mathematically approximated by a pin-hole camera model. In this model, no lens or sensor distortions are assumed. A target point in the image space, the homologous target point in the object space, and the hypothetical perspective centre of the camera are assumed to be collinear. This allows well-established camera registration techniques (such as the photogrammetric bundle adjustment) to be applied, which relate multiple radiographs with the calibration target field at various positions and orientations (Equation 1).

$$p_{ij} = q_j(P_i - T_j)q_j^c \qquad (1)$$

where, $p_{ij} = [x_{ij} - x_p - \Delta x, y_{ij} - y_p - \Delta y, -c]^T$ is the image measurement coordinates of target $i$ in exposure $j$
$P_i = [X_i, Y_i, Z_i]^T$ is the object space coordinates of target $i$ on the phantom
$T_j = [X_{oj}, Y_{oj}, Z_{oj}]^T$ is the position of the X-ray system relative to the phantom in exposure $j$
$q_j$ = unit quaternion representing rotation of the X-ray system relative to the phantom in exposure $j$. Superscript 'c' represents the quaternion conjugate.

This is an idealized model that requires a minimum of three non-collinear targets to solve. In practice, many more targets are observed and statistical inference techniques such as Maximum Likelihood Estimation (MLE) are used to obtain the best estimate of the unknown interior orientation parameters (IOP = [$x_p, y_p, c$]), exterior orientation parameters (EOP = [$T_j, q_j$]), and object space target coordinates. With a high redundancy, additional parameters (AP = [$\Delta x, \Delta y$]) can be included in the bundle adjustment to model any systematic errors in the device.

Selecting the proper model complexity to estimate the systematic errors requires a delicate balance between bias and variance. Having too many AP will result in an over-fitting problem, while not including enough AP will end up with a high bias. Previously, Lichti et al. (2015) proposed a systematic error model with up to 31 parameters. To choose the optimal number of parameters is a lengthy process: an experienced photogrammetrist will need to perform the bundle adjustment with self-calibration many times, starting with a simple model (e.g. no AP) and gradually increasing the model complexity (e.g. adding one AP to the adjustment at a time). With each bundle adjustment, graphical and statistical analysis is performed to assess the added value of the new parameter. If the new parameter reduces the root-mean-squared-error (RMSE) of the observation residuals, is found to be statistically significant using the student t-test, and reduces any systematic trends visually seen in the residual plots then that AP is deemed relevant in the model. In cases where not all the above conditions are satisfied, an expert's judgement is required to determine if that AP should be added. Even for a well-trained photogrammetrist this can become a labour-intensive process since there is such a large number of potential AP to choose from. For a non-expert, the upper limit of possible models to select can be estimated using the following expression:

$$\sum_{n=0}^{31} C(31, n) = \sum_{n=0}^{31} \frac{31!}{n!(31-n)!} \approx 2 \times 10^9$$

To automate this model selection process and to make it operator independent, a k-nearest-neighbour (kNN) regression approach is used to model the systematic errors in the residuals. This data-driven machine learning approach assumes that spatially nearby residuals are correlated, and therefore can be approximated by averaging the k nearest residuals. It is considered a parameter-free approach (as no parameters need to be learned using least-squares estimation) and can be highly scalable to large amounts of data if a KD-tree structure is used for organizing the data since the residuals are only in 2D. However, one hyperparameter (k) still needs to be tuned, which indirectly defines the neighbourhood size used in the regression. To tune the k parameter using the data itself a 10-fold cross-validation was used. The weighted L2-norm is used as the error metric (Equation 2) with the grid search approach to find the optimal k.

$$G = (\vec{r} - \vec{g}(x,y))^T C_r^{-1} (\vec{r} - \vec{g}(x,y)) \qquad (2)$$

where, $\vec{r}$ is the vector of residuals
$C_r$ is the covariance matrix of the residuals
$x$ and $y$ are the cartesian coordinates in image space
$\vec{g}(x,y)$ is the vector of predicted residuals

### 3.1 Proposed Methodology

The X-ray calibration approach adopted in this paper draws on the concept of grey-box system identification. By initializing $\Delta x$ and $\Delta y$ to be zero, a robust photogrammetric bundle adjustment is first performed by minimizing the negative logarithm of the student-t probability distribution (Equation 3). This is equivalent to finding the point of maximum likelihood. Since the collinearity condition is non-linear, the model is linearized using a first-order Taylor series expansion and the unknown parameters are updated iteratively. At every iteration, the step is calculated using the popular trust-region method, the Levenberg–Marquardt algorithm. Once the MLE has converged, the residuals and corresponding variance-covariance matrix are computed. This then serves as the input to the second step, which is the kNN regression.

$$F = \max_{\vec{\theta}} \frac{\Gamma\left(\frac{v}{2} + \frac{D}{2}\right)}{\Gamma\left(\frac{v}{2}\right)} \frac{|C_l|^{-\frac{1}{2}}}{(v\pi)^{\frac{D}{2}}} \left[1 + \frac{\left(\vec{l} - f(\vec{\theta})\right)^T C_l^{-1} \left(\vec{l} - f(\vec{\theta})\right)}{v}\right]^{-\frac{v}{2} - \frac{D}{2}} \qquad (3)$$

where, $\vec{l}$ is the vector of image measurements
$C_l$ is the covariance matrix of the observations
$\vec{\theta}$ is the vector of unknown parameters
$f(\vec{\theta})$ represents the collinearity condition
$v$ is the degrees-of-freedom
$D$ is the number of observations

During the kNN regression, the best estimate of $\Delta AP$ is determined after automatically tuning the hyperparameter $k$ using cross-validation. Only the residual data that were considered inliers from the bundle adjustment are used for training the regressor. The $\Delta x$ and $\Delta y$ are then updated by the $\Delta AP$ predicted by the regressor ($AP_{next} = AP_{previous} + \Delta AP$). The process of performing a robust photogrammetric bundle adjustment step followed by the kNN regression is then repeated until convergence. This iterative self-calibration adjustment converges when both the weighted cost function of the bundle adjustment (i.e. $F$) and the kNN regression (i.e. $G$) are minimized.

After convergence, not only can the IOP, EOP, and object space coordinates be obtained together with their standard deviations, a kNN regressor that has learned the irregularly-spaced systematic error corrections of the residuals is also available. For new radiographs that are captured, the distortion correction at every pixel location can be predicted by the kNN regressor.

## 4. EXPERIMENTATION

The same data as in Lichti et al. (2015) were used, where a three-dimensional cubic target frame (i.e. a phantom with 503 targets) was imaged using two fluoroscopic image systems simultaneously. Each fluoroscopic system consists of an X-ray source, an image intensifier with a fluorescent screen, and a high-speed solid-state optical camera. The fluoroscopic system was static during the entire experiment while the phantom was repositioned with various orientations within the volume of interest with the help of a height-adjustable turntable. A total of 150 image frames per fluoroscopic system was processed using bundle adjustment with self-calibration by an expert photogrammetrist. The reconstructed 3D object space and the resected virtual camera locations using all images serve as the reference solution. A subset of this data, i.e. 15 of the 150 images were uniformly sampled from each fluoroscopic system and processed using the proposed algorithm. This subset acted as the training data; the remaining 135 images were used as testing data. It was hypothesized that if the recovered systematic distortion profile using 10% of the data would be comparable to the result from using 150 images with an expert's judgement on model selection, then the proposed method has the potential to further automate the calibration process without compromising the quality of the calibration solution.

The proposed calibration method (i.e. extending conventional bundle adjustment with machine learning) can be approached in various ways. For example, there is the preconception that machine learning can learn everything if given sufficient data, even depth (Sinz et al., 2004). While this may be true, incorporating prior knowledge about the problem can strengthen the solution. Thus, for the experiments described previously, the calibration was done in two ways: (1) the kNN regressor was used to learn both AP and IOP, and (2) the kNN regressor was used to learn the AP only (with the IOP being modelled parametrically). Unlike the AP, the IOP are expected to be present in all imaging systems and their mathematical form is known. The IOP consist of merely three unknown parameters, and they are comparable to a bias and scale factor. By solving for the IOP using the standard parametric form in the bundle adjustment, it is expected that the solution can be improved. If this is true, then a similar argument can be made about estimating the EOP and object space coordinates using the parametric form rather than learning it from data.

## 5. RESULTS AND ANALYSES

### 5.1 Efficacy of Proposed Calibration Method

The proposed calibration method divides up the numerical optimization process into two steps. In the first step the reprojection errors are minimized using a robust bundle adjustment. A surface is then fitted to the residuals using kNN regression. Figure 1 shows the monotonic progressive reduction in the quadratic costs. The gradient begins to diminish around 20 iterations. The weighted average of the total cost (i.e. $F + G$) shown in Figure 2 demonstrated that a stable local minimum can be found after around 30 iterations by following the gradient.

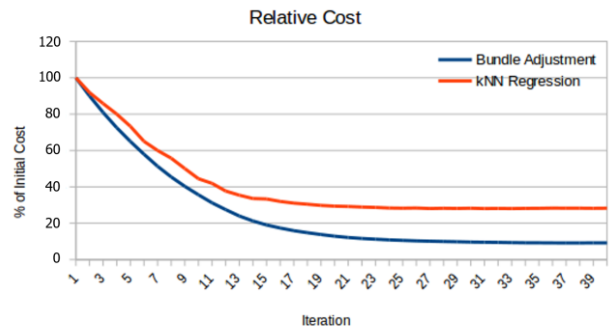

Figure 1: Relative cost at every iteration expressed as a percentage of the initial cost. $F$ is shown in blue and $G$ is shown in red.

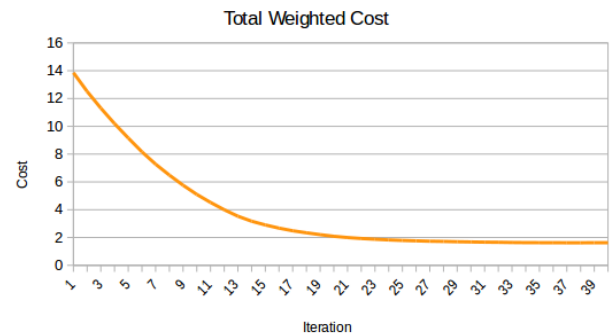

Figure 2: Weighted average cost of both the bundle adjustment and kNN regression

### 5.2 Error Modelling of Fluoroscopic Imaging System 1

It is hypothesized that in the absence of systematic errors, the distribution of residuals will follow a Gaussian probability distribution. Figures 3 and 4 show the histogram of the normalized image residuals in x and y, respectively. It can be seen that the distribution resembles a bell shape much better after modelling for the AP. In addition, by explicitly modelling for the IOP rather than learning it from the data showed slight improvements in the reprojection errors.

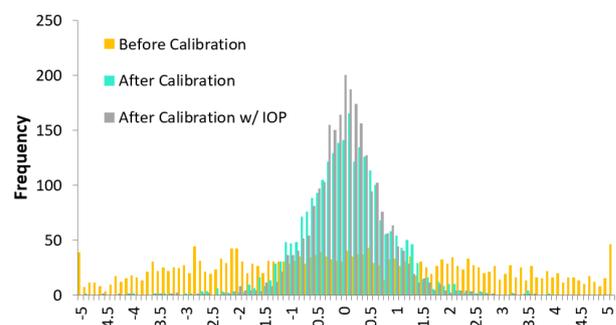

Figure 3: Histogram of the x-image residuals in fluoroscopic system 1

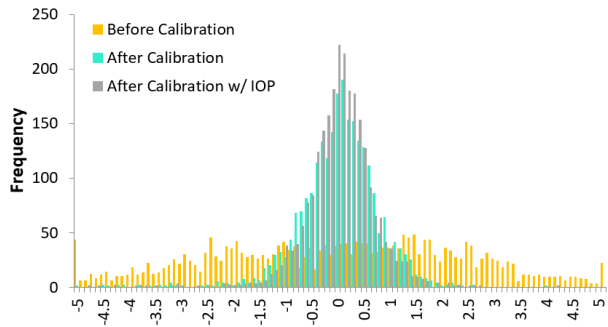

Figure 4: Histogram of the y-image residuals in fluoroscopic system 1

To analyze the correlation between the reprojection errors and the spatial distribution, the residuals in x and y are plotted as a function of their x and y pixel locations in the image (Figures 5 and 6). The residuals after calibration appear to have a smaller spread overall, with the residuals being slightly larger near the peripheral of the radiograph. This is expected in kNN regression because of the lack of data points near edges of the image, which results in all the data points being on one side of the query point. In other words, near the edge of the radiograph, kNN regression behaves more like an extrapolator than an interpolator.

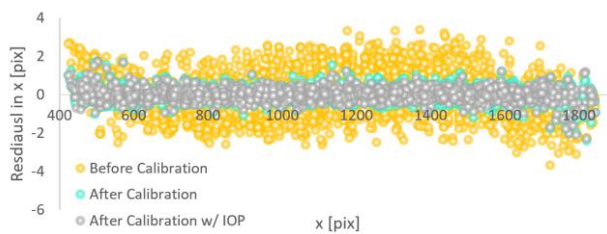

Figure 5: x-image residuals as a function of their column number in the image

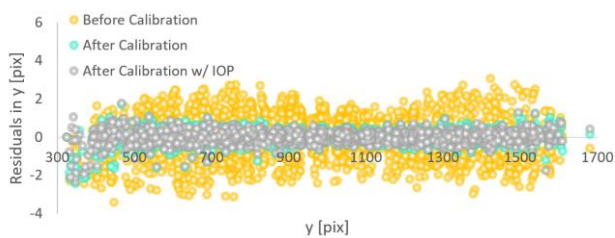

Figure 6: y-image residuals as a function of their row number in the image

Table 1 summarizes the reprojection errors. It can be observed that the fluoroscopic imaging data obeys the pin-hole camera model with precision of approximately one pixel. After calibrating the X-ray system a significant reduction in the RMSE can be obtained (i.e. from pixel-level to sub-pixel-level).

| | Cost | RMSE [pix] | | Improvement [%] | | |
|---|---|---|---|---|---|---|
| | $r^T C_r^{-1} r$ | $x$ | $y$ | $r^T C_r^{-1} r$ | $x$ | $y$ |
| Before | 4.72E+04 | 1.148 | 1.070 | N/A | N/A | N/A |
| After | 3.47E+03 | 0.346 | 0.339 | 92.638 | 69.839 | 68.372 |
| After w/ IOP | 2.48E+03 | 0.303 | 0.283 | 94.746 | 73.624 | 73.536 |

Table 1: Reprojection errors of fluoroscopic system 1

Studying the reprojection errors in image space is useful because it is the objective function that is being minimized during MLE. However, as a clinician or technician, it is the object space accuracy improvement that is of utmost interest. Tables 2 and 3 show the measurement errors in the 3D object space (X, Y, and Z) as well as the errors in the estimated X-ray sensor location relative to the phantom ($X_o$, $Y_o$, $Z_o$). For the in-sample errors, the data-driven error compensation model was able to reduce the object space reconstruction errors by about 80%. Since the X-ray system has a long principal distance, the object space can be measured more accurately than the camera location (Stamatopoulos et al., 2010). Nevertheless, the learned calibration model using kNN improved the estimation of the sensor positions. The estimated sensor position is significantly improved when the IOP are modelled in the bundle adjustment. This is similar to removing the bias and scale from the data in which the kNN regression is performed. The large improvements in the estimated network station geometry (centimetre-level to millimetre-level) gave rise to a further improvement of the object space measurement accuracy (~5%).

Since the in-sample errors (Table 2) are more sensitive to projective compensation, the out-of-sample errors (Table 3) are also reported. These data had never been seen by the bundle adjustment or kNN regressor, therefore providing a more realistic assessment of the actual quality of the calibration parameters when they are used to correct images. A pattern similar to the in-sample errors can be observed in the out-of-sample errors: the calibrated radiographs with the IOP included in the collinearity conditions provided the best object space accuracy. Even though the IOP can be learned from the data using kNN, modelling them using explicit parameters in the bundle adjustment yielded superior results.

| | RMSE [mm] | | | Improvement [%] | |
|---|---|---|---|---|---|
| | Before | After | After w/ IOP | After | After w/ IOP |
| $X$ | 1.271 | 0.223 | 0.148 | 82.478 | 88.339 |
| $Y$ | 0.848 | 0.179 | 0.133 | 78.852 | 84.269 |
| $Z$ | 1.384 | 0.215 | 0.153 | 84.476 | 88.976 |
| $X_o$ | 18.382 | 14.773 | 3.179 | 19.637 | 82.704 |
| $Y_o$ | 27.372 | 14.565 | 3.280 | 46.788 | 88.018 |
| $Z_o$ | 16.2599 | 15.477 | 3.421 | 4.815 | 78.960 |

Table 2: In-sample errors of fluoroscopic system 1

| | RMSE [mm] | | | Improvement [%] | |
|---|---|---|---|---|---|
| | Before | After | After w/ IOP | After | After w/ IOP |
| $X$ | 0.938 | 0.152 | 0.069 | 83.771 | 92.687 |
| $Y$ | 0.509 | 0.111 | 0.047 | 78.192 | 90.749 |
| $Z$ | 0.926 | 0.162 | 0.072 | 82.483 | 92.196 |
| $X_o$ | 15.343 | 14.721 | 3.140 | 4.052 | 79.532 |
| $Y_o$ | 23.221 | 15.499 | 3.485 | 33.254 | 84.990 |
| $Z_o$ | 16.222 | 15.065 | 3.285 | 7.129 | 79.746 |

Table 3: Out-of-sample errors of fluoroscopic system 1

### 5.3 Error Modelling of Fluoroscopic Imaging System 2

To further assess the performance and behaviour of the proposed calibration method, the previous assessments were repeated on the second X-ray fluoroscopic system. The trends are similar to the calibration of fluoroscopic system 1. The residuals more closely follow a normal distribution post-calibration, with the bell curve being narrower when the IOP are estimated in the bundle adjustment rather than being learned in the kNN regression (Figure 7, 8, 9, 10). The overall reprojection errors reduced from greater than a pixel to less than half a pixel after error modelling (Table 4).

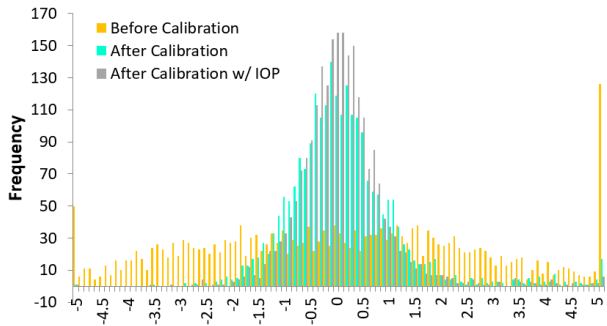

Figure 7: Histogram of the x-image residuals in fluoroscopic system 2

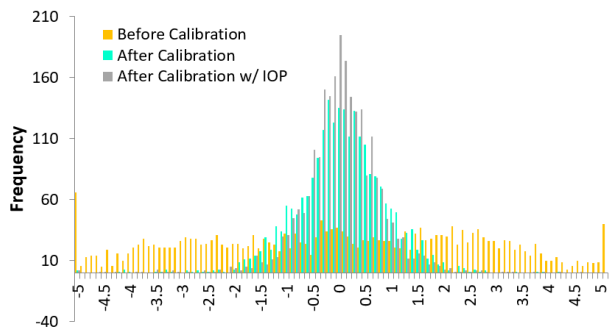

Figure 8: Histogram of the y-image residuals in fluoroscopic system 2

| | Cost | RMSE [pix] | | Improvement [%] | | |
|---|---|---|---|---|---|---|
| | $r^T C_r^{-1} r$ | $x$ | $y$ | $r^T C_r^{-1} r$ | $x$ | $y$ |
| Before | 4.62E+04 | 1.235 | 1.203 | N/A | N/A | N/A |
| After | 6.13E+03 | 0.505 | 0.368 | 86.722 | 59.074 | 69.399 |
| After w/ IOP | 3.78E+03 | 0.384 | 0.323 | 91.811 | 68.870 | 73.182 |

Table 4: Reprojection errors of fluoroscopic system 2

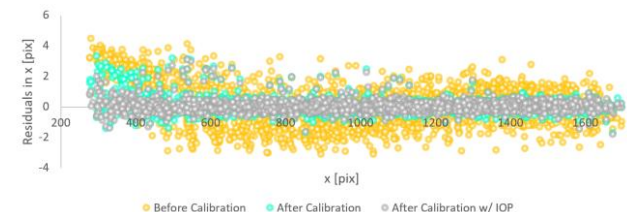

Figure 9: x-image residuals as a function of their column number in the image

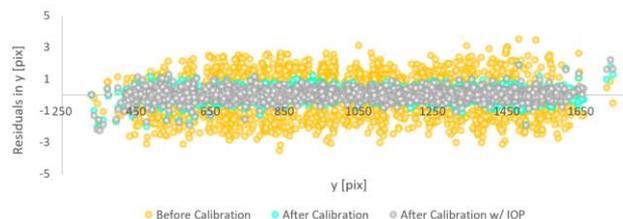

Figure 10: y-image residuals as a function of their row number in the image

The object space measurement accuracy of system 2 is slightly worse and less uniform than system 1; this might be due to the stability and manufacturing of the system. Even though both systems used identical components there can still be small variations in their build. Regardless, the proposed calibration method is able to improve the object space reconstruction accuracy both in-sample and out-of-sample for system 2 (Tables 5 and 6).

When looking at the out-of-sample errors, it was noticed that the estimated sensor position errors were greater with the error correction model when the IOP were learned by the kNN regression. In this dataset, solving for the IOP explicitly in the bundle adjustment has a small improvement to the object space reconstruction quality and was able to reduce the sensor position errors rather than increasing it.

| | RMSE [mm] | | | Improvement [%] | |
|---|---|---|---|---|---|
| | Before | After | After w/ IOP | After | After w/ IOP |
| $X$ | 1.973 | 0.450 | 0.405 | 77.188 | 79.461 |
| $Y$ | 0.941 | 0.237 | 0.208 | 74.815 | 77.875 |
| $Z$ | 1.295 | 0.467 | 0.434 | 63.968 | 66.504 |
| $X_o$ | 20.575 | 19.311 | 2.381 | 6.146 | 88.427 |
| $Y_o$ | 20.121 | 19.749 | 2.431 | 1.847 | 87.916 |
| $Z_o$ | 21.536 | 18.339 | 2.300 | 14.844 | 89.320 |

Table 5: In-sample errors of fluoroscopic system 2

| | RMSE [mm] | | | Improvement [%] | |
|---|---|---|---|---|---|
| | Before | After | After w/ IOP | After | After w/ IOP |
| $X$ | 0.669 | 0.150 | 0.121 | 77.561 | 81.941 |
| $Y$ | 0.510 | 0.099 | 0.055 | 80.540 | 89.232 |
| $Z$ | 0.708 | 0.131 | 0.096 | 81.506 | 86.439 |
| $X_o$ | 19.298 | 19.138 | 4.663 | 0.830 | 75.835 |
| $Y_o$ | 16.010 | 18.966 | 1.825 | -18.468 | 88.602 |
| $Z_o$ | 16.984 | 18.845 | 1.939 | -10.960 | 88.583 |

Table 6: Out-of-sample errors of fluoroscopic system 2

### 5.4 Simultaneous Calibration of Two Fluoroscopic Imaging Systems

In clinics with a dual-fluoroscopic imaging system for tracking 3D motions, both X-ray systems can be calibrated simultaneously. The image measurements in each radiograph are independent, but if they are observing the same phantom then they become correlated through the object space coordinates. By performing bundle adjustment for the two fluoroscopic imaging systems jointly and training a kNN regressor for each system, the following image space errors are reported (Table 7).

| | Cost | RMSE [pix] | | Improvement [%] | | |
|---|---|---|---|---|---|---|
| | $r^T C_r^{-1} r$ | $x$ | $y$ | $r^T C_r^{-1} r$ | $x$ | $y$ |
| Before | 1.139E+05 | 1.307 | 1.152 | N/A | N/A | N/A |
| After | 1.039E+04 | 0.451 | 0.363 | 90.878 | 65.474 | 68.508 |
| After w/ IOP | 6.805E+03 | 0.358 | 0.314 | 94.027 | 72.623 | 72.724 |

Table 7: Reprojection errors of fluoroscopic systems 1 and 2

The in-sample and out-of-sample object space errors (Tables 8 and 9) are comparable to the cases where each fluoroscopic system was calibrated independently (Tables 2, 3, 5, and 6). Only sharing the same object space targets does not seem to provide any statistically significant benefits to the object space reconstruction quality. Therefore, it can be argued that not calibrating both X-ray systems together is preferred because of the reduced computation load. To experience benefits of object space reconstruction accuracy in a dual-fluoroscopic image system, the two X-ray systems should be rigidly mounted together so that a relative position and orientation constraint can be enforced in the MLE (Lichti et al., 2015).

|   | RMSE [mm] | | | Improvement [%] | |
|---|---|---|---|---|---|
|   | Before | After | After w/ IOP | After | After w/ IOP |
| $X$ | 1.161 | 0.369 | 0.315 | 68.201 | 72.871 |
| $Y$ | 1.046 | 0.238 | 0.179 | 77.276 | 82.918 |
| $Z$ | 1.025 | 0.291 | 0.215 | 71.603 | 78.990 |
| $X_o$ | 18.234 | 16.583 | 2.640 | 9.056 | 85.520 |
| $Y_o$ | 20.844 | 16.881 | 2.717 | 19.012 | 86.963 |
| $Z_o$ | 17.120 | 16.466 | 2.655 | 3.821 | 84.490 |

Table 8: In-sample errors of fluoroscopic system 1 and 2

|   | RMSE [mm] | | | Improvement [%] | |
|---|---|---|---|---|---|
|   | Before | After | After w/ IOP | After | After w/ IOP |
| $X$ | 0.674 | 0.135 | 0.095 | 79.951 | 85.916 |
| $Y$ | 0.465 | 0.093 | 0.039 | 79.993 | 91.671 |
| $Z$ | 0.683 | 0.141 | 0.091 | 79.401 | 86.693 |
| $X_o$ | 17.043 | 16.486 | 3.407 | 3.265 | 80.011 |
| $Y_o$ | 19.608 | 16.806 | 2.528 | 14.290 | 87.106 |
| $Z_o$ | 16.221 | 16.550 | 2.524 | -2.027 | 84.441 |

Table 9: Out-of-sample errors of fluoroscopic system 1 and 2

## 6. CONCLUSION AND FUTURE WORK

Fluoroscopic imaging systems allow the use of low dose radiation to look under the skin of humans non-invasively for clinical diagnoses. This modality is particularly valuable in studying dynamic data because it has a high frame rate. To use this system for quantitative analysis and to improve the geometric accuracy of the images for qualitative assessments, systematic errors of the complete system need to be removed. Previous research has already demonstrated that using photogrammetric bundle adjustment to do a software calibration of the imaging system can improve both the precision and accuracy of the system. This paper presented an extension by adding a machine learning approach using kNN regression to automate the model selection process in the bundle adjustment, thus making it easier for a non-expert to perform the calibration. It has been shown in this paper that not only can the proposed data-driven method make the calibration process less operator dependent, it was able to achieve a similar level of accuracy as the parameter-driven approach.

While all information can be learned from the data using machine learning – including depth and camera pose – it was shown that if a geometric relationship is expected to exist in the system, it is more effective to model them explicitly using well-established parametric models. For example, it was found that when both the IOP and AP are learned using kNN regression, the precision and accuracy (using the same input dataset) are both lower than if the IOP are modelled using the conventional approach in the bundle adjustment.

At present, most hospitals only have single fluoroscopic systems available, and therefore the proposed method is applicable. Future work will investigate using relative orientation constraints with this method for dual-fluoroscopic imaging systems that have a fixed baseline. It was found in this paper that even if two fluoroscopic imaging systems were calibrated simultaneously and they share the same object space targets, the calibration result is very similar to the scenario where the two fluoroscopic systems were calibrated separately. Hence, there is little benefit to perform a multi-system calibration when a relative orientation constraint cannot be enforced.

## ACKNOWLEDGEMENTS

The colour scheme for the figures is inspired by the International JK Conference 2018 held in Calgary, Canada.

## REFERENCES

Al-Durgham, K., Lichti, D., Kuntze, G., Sharma, G., & Ronsky, J. (2016). Toward an automatic calibration of dual fluoroscopy imaging systems. *International Archives of Photogrammetry, Remote Sensing and Spatial Information Sciences, XLI-B5*, 757-764.

Lichti, D., Sharma, G., Kuntze, G., Mund, B., Beveridge, J., & Ronsky, J. (2015). Rigorous geometric self-calibrating bundle adjustment for a dual fluoroscopic imaging system. *IEEE Transaction on Medical Imaging, 34*(2), 589-598.

Sinz, F., Candela, J., Bakır, G., Rasmussen, C., & Franz, M. (2004). Learning depth from stereo. In C. Rasmussen, H. Bülthoff, B. Schölkopf, & M. Giese, *Pattern Recognition. Lecture Notes in Computer Science* (Vol. 3175, pp. 245-252). Springer, Berlin, Heidelberg.

Stamatopoulos, C., Fraser, C., & Cronk, S. (2010). On the self-calibration of long focal length lenses. *International Archives of Photogrammetry, Remote Sensing and Spatial Information Sciences, Vol. XXXVIII, Part 5*, 560-564.

You, B., Siy, P., Anderst, W., & Tashman, S. (2001). Vivo measurement of 3-D skeletal kinematics from sequences of biplane radiographs: Applicaiton to knee kinematics. *IEEE Transaction on Medical Imaging, 20*(6), 514-525.